# Critical thermodynamics of two-dimensional $N$-vector cubic model in the five-loop approximation

P.Calabrese[*][1], E.V.Orlov[2], D.V.Pakhnin[2], A.I.Sokolov[†][2]

[1] Scuola Normale Superiore and INFN,
Piazza dei Cavalieri 7, I–56126 Pisa, Italy

[2] Department of Physical Electronics,
Saint Petersburg Electrotechnical University,
Professor Popov Street 5, St. Petersburg 197376, Russia

The critical behavior of the two-dimensional $N$-vector cubic model is studied within the field-theoretical renormalization-group (RG) approach. The $\beta$ functions and critical exponents are calculated in the five-loop approximation, RG series obtained are resummed using Padé-Borel-Leroy and conformal mapping techniques. It is found that for $N = 2$ the continuous line of fixed points is well reproduced by the resummed RG series and an account for the five-loop terms makes the lines of zeros of both $\beta$ functions closer to each other. For $N \geqslant 3$ the five-loop contributions are shown to shift the cubic fixed point, given by the four-loop approximation, towards the Ising fixed point. This confirms the idea that the existence of the cubic fixed point in two dimensions under $N > 2$ is an artifact of the perturbative analysis. In the case $N = 0$ the results obtained are compatible with the conclusion that the impure critical behavior is controlled by the Ising fixed point.

## 1. Introduction

The two-dimensional (2D) model with $N$-vector order parameter and cubic anisotropy is known to have a rich phase diagram; it contains, under different values of $N$ and of the anisotropy parameter, the Ising-like and Kosterlitz-Thouless critical points, lines of the first-order phase transitions, and the line of the second-order transitions with continuously varying critical exponents (see, e. g. [1–3] for review).

[*]E-mail: calabres@df.unipi.it
[†]E-mail: ais2002@mail.ru

This model is related to many other familiar models in various particular cases, while for $N \to 0$ it describes the critical behavior of 2D weakly disordered Ising systems. Moreover, exact solutions are known for the 2D cubic model in the several limits such as an Ising decoupled limit, the limit of extremely strong anisotropy for $N > 2$ [2,4] and the replica limit $N \to 0$ [2,5,6]. The mappings, in particular regions of the phase diagram, with the $N$-color Ashkin-Teller models, discrete cubic models, and planar model with fourth order anisotropy give further information about the critical behavior. All these issues are reviewed in [3] and we will not repeat here. These features make the 2D $N$-vector cubic model a convenient and, perhaps, unique testbed for evaluation of the analytical and numerical power of perturbative methods widely used nowadays in the theory of critical phenomena. The field-theoretical renormalization-group (RG) approach in physical dimensions is among of them.

Recently, the critical behavior of the 2D $N$-vector cubic model was explored using the renormalization-group technique in the space of fixed dimensionality [3]. The four-loop expansions for the $\beta$-functions and critical exponents were calculated and analyzed using the Borel transformation combined with the conformal mapping and Padé-approximant techniques as a tool for resummation of the divergent RG series. The most part of predictions obtained within the renormalization group approach turned out to be in accord with the known exact results. At the same time, some findings were quite new. In particular, for $N > 2$ the resummed four-loop RG expansions for $\beta$-functions were found to yield a cubic fixed point with (almost) marginal stability; this point does not correspond to any of the critical asymptotes revealed by exact methods ever applied. Although the stability properties of the cubic fixed point look very similar to those of its Ising counterpart, these points were found to lie too far from each other (for moderate $N$) to consider the distance between them as a splitting caused by the limited accuracy of the RG approximation employed.

It is worth noting that this situation is quite different from what we have in three dimensions. Indeed, for the 3D cubic model the structure of the RG flow diagram is known today with a rather high accuracy. Recent five-loop [7–9] and six-loop [10,11] RG calculations certainly confirmed that for $N > 2$ the cubic fixed point does not merge with any other fixed point and governs the specific anisotropic mode of critical behavior, distinguishable from the Ising and Heisenberg modes (see, e. g. [12]).

It is very desirable, therefore, to clear up to what extent the location of the cubic fixed point in two dimensions is sensitive to the order of the RG calculations and, more generally, whether this point really exists at the flow diagram or its appearance is the approximation artifact caused by the finiteness of the perturbative series and by an ignorance of the confluent singularities significant in two dimensions [13–16].

Of prime interest is also the situation with the line of fixed points that should run, under $N = 2$, from the Ising fixed point to the Heisenberg one. Within the four-loop approximation, the zeros of $\beta$-functions for the $O(N)$-symmetric and anisotropic coupling constants form two lines that for $N = 2$ are practically parallel to each other and separated by the distance that is smaller than the error bar appropriate to the working approximation [3]. Will the higher-order contributions keep these two lines parallel? Will an account for the higher-order terms further diminish the distance between these lines or their splitting should be attributed, at least partially, to the effect of the singular terms just mentioned?

To answer the above questions, it is necessary to analyze the critical behavior of the 2D cubic model in the higher perturbative orders. Recently, the renormalization-group expansions for the 2D $O(N)$-symmetric model were obtained within the five-loop approximation [15]. In the course of this study, all the integrals corresponding to the five-loop four-leg and two-leg Feynman graphs have been evaluated. This makes it possible to investigate the critical thermodynamics of anisotropic 2D models with several couplings in the five-loop approximation. In this paper, such an investigation will be carried out for the 2D $N$-vector model with cubic anisotropy.

## 2. Renormalization group expansions

In order to study the effect of cubic anisotropies one usually considers the $\phi^4$ theory [17,18]:

$$\mathcal{H} = \int \mathrm{d}^d x \left\{ \frac{1}{2} \sum_{i=1}^{N} \left[ (\partial_\mu \phi_i)^2 + r \phi_i^2 \right] + \frac{1}{4!} \sum_{i,j=1}^{N} \left( u_0 + v_0 \delta_{ij} \right) \phi_i^2 \phi_j^2 \right\}, \qquad (1)$$

in which the added cubic term breaks explicitly the $O(N)$ invariance leaving a residual discrete cubic symmetry given by the reflections and permutations of the field components. In two dimensions the effect of anisotropy is particularly important: systems possessing continuous symmetry do not exhibit conventional long-range order at finite temperature, while models with discrete symmetry do undergo phase transitions into conventionally ordered phase.

In general, the model (1) has four fixed points: the trivial Gaussian one, the Ising one in which the $N$ components of the field decouple, the O($N$)-symmetric and the cubic fixed points. The Gaussian fixed point is always unstable, and so is the Ising fixed point for $d > 2$ [17]. Indeed, in the latter case, it is natural to interpret equation (1) as the Hamiltonian of $N$ Ising-like systems coupled by the $O(N)$-symmetric term. But this interaction is the sum of the products of the energy operators of the different Ising systems. Therefore, at the Ising fixed point, the crossover exponent associated with the $O(N)$-symmetric quartic term should be given by the specific-heat critical exponent $\alpha_\mathrm{I}$ of the Ising model, independently of $N$. Since $\alpha_\mathrm{I}$ is positive for all $d > 2$ the Ising fixed point is unstable. Obviously, in two dimensions this argument only told us that the crossover exponent at this fixed point vanishes. Higher order corrections to RG equation may lead either to a marginally stable fixed point or to a line of fixed points. It was argued that for $N \geqslant 3$ the first possibility is realized, while for $N = 2$ the second one holds (see [1–3] and references therein).

The stability properties of the O($N$)-symmetric and of the cubic fixed points depend on $N$. For sufficiently small values of $N$, $N < N_\mathrm{c}$, the O($N$)-symmetric fixed point is stable and the cubic one is unstable. For $N > N_{\rm c}$, the opposite is true: the renormalization-group flow is driven towards the cubic fixed point, which now describes the generic critical behavior of the system. At $N = N_{\rm c}$, the two fixed points should coincide for $d > 2$. At $d = 2$, it is expected that $N_{\rm c} = 2$ and a line of fixed points connecting the Ising and the $O(2)$-symmetric fixed points exists [1,3].

The fixed-dimension field-theoretical approach is known to be a powerful tool in studying the critical properties of three-dimensional systems belonging to the $O(N)$ and more complicated universality classes (see, e.g., [18–20]). In this approach one performs an expansion in powers of appropriately defined zero-momentum quartic couplings and renormalizes the theory by a set of zero-momentum conditions for the (one-particle irreducible) two-point and four-point correlation functions:

$$\Gamma^{(2)}_{ab}(p) = \delta_{ab} Z_\phi^{-1} \left[ m^2 + p^2 + O(p^4) \right], \tag{2}$$

$$\Gamma^{(4)}_{abcd}(0) = Z_\phi^{-2} m^2 \left[ \frac{u}{3} \left( \delta_{ab}\delta_{cd} + \delta_{ac}\delta_{bd} + \delta_{ad}\delta_{bc} \right) + v\, \delta_{ab}\delta_{ac}\delta_{ad} \right]. \tag{3}$$

They relate the inverse correlation length (mass) $m$ and the zero-momentum quartic couplings $u$ and $v$ to the corresponding Hamiltonian parameters $r$, $u_0$, and $v_0$:

$$u_0 = m^2 u Z_u Z_\phi^{-2}, \qquad v_0 = m^2 v Z_v Z_\phi^{-2}. \tag{4}$$

In addition, one introduces the function $Z_t$ defined by the relation

$$\Gamma^{(1,2)}_{ab}(0) = \delta_{ab} Z_t^{-1}, \tag{5}$$

where $\Gamma^{(1,2)}$ is the (one-particle irreducible) two-point function with an insertion of $\phi^2/2$.

From the pertubative expansions of the correlation functions $\Gamma^{(2)}$, $\Gamma^{(4)}$, and $\Gamma^{(1,2)}$ and the above relations, one derives the functions $Z_\phi(u,v)$, $Z_u(u,v)$, $Z_v(u,v)$, and $Z_t(u,v)$ as double expansions in $u$ and $v$.

The fixed points of the theory are given by the common zeros of the $\beta$-functions

$$\beta_u(u,v) = m \left. \frac{\partial u}{\partial m} \right|_{u_0,v_0}, \qquad \beta_v(u,v) = m \left. \frac{\partial v}{\partial m} \right|_{u_0,v_0}. \tag{6}$$

The stability properties of the fixed points are controlled by the eigenvalues $\omega_i$ of the matrix

$$\Omega = \begin{pmatrix} \dfrac{\partial \beta_u(u,v)}{\partial u} & \dfrac{\partial \beta_u(u,v)}{\partial v} \\ \dfrac{\partial \beta_v(u,v)}{\partial u} & \dfrac{\partial \beta_v(u,v)}{\partial v} \end{pmatrix}, \tag{7}$$

computed at the given fixed point: a fixed point is stable if both eigenvalues are positive. The eigenvalues $\omega_i$ are related to the leading scaling corrections, which vanish as $\xi^{-\omega_i} \sim |t|^{\Delta_i}$ where $\Delta_i = \nu \omega_i$.

One also introduces the functions

$$\eta_{\phi,t}(u,v) = \left. \frac{\partial \ln Z_{\phi,t}}{\partial \ln m} \right|_{u_0,v_0} = \beta_u \frac{\partial \ln Z_{\phi,t}}{\partial u} + \beta_v \frac{\partial \ln Z_{\phi,t}}{\partial v}, \tag{8}$$

so that the critical exponents are obtained from

$$\eta = \eta_\phi(u^*, v^*), \tag{9}$$
$$\nu = [2 - \eta_\phi(u^*, v^*) + \eta_t(u^*, v^*)]^{-1}, \tag{10}$$
$$\gamma = \nu(2 - \eta), \tag{11}$$

where $(u^*, v^*)$ is the position of the stable fixed point.

Here, we present the perturbative expansions of the RG functions (6) and (8) up to five loops. The results are written in terms of the rescaled couplings

$$u \equiv \frac{8\pi}{3} R_N \, \bar{u}, \qquad v \equiv \frac{8\pi}{3} \, \bar{v}, \tag{12}$$

where $R_N = 9/(8 + N)$.

**Table 1.** The coefficients $b_{ij}^{(u)}$, cf. equation (13).

| $i, j$ | $(N+8)^i b_{ij}^{(u)}$ |
|---|---|
| 2,0 | $-47.6751 - 10.335\,N$ |
| 1,1 | $-8.39029$ |
| 0,2 | $-0.21608$ |
| 3,0 | $524.377 + 149.152\,N + 5.00028\,N^2$ |
| 2,1 | $144.813 + 7.27755\,N$ |
| 1,2 | $10.0109 + 0.0583278\,N$ |
| 0,3 | $0.231566$ |
| 4,0 | $-7591.11 - 2611.15\,N - 179.698\,N^2 - 0.088843\,N^3$ |
| 3,1 | $-2872.08 - 291.254\,N + 0.126813\,N^2$ |
| 2,2 | $-330.599 - 5.97086\,N$ |
| 1,3 | $-16.0559 - 0.0578955\,N$ |
| 0,4 | $-0.311695$ |
| 5,0 | $133972. + 53218.6\,N + 5253.56\,N^2 + 80.3097\,N^3 - 0.0040796\,N^4$ |
| 4,1 | $64819.7 + 9554.31\,N + 164.916\,N^2 + 0.145241\,N^3$ |
| 3,2 | $10584.2 + 439.25\,N + 1.29693\,N^2$ |
| 2,3 | $818.21 + 8.3695\,N$ |
| 1,4 | $32.7458 + 0.0796603\,N$ |
| 0,5 | $0.555161$ |

The resulting series are

$$\bar{\beta}_{\bar{u}} = -\bar{u} + \bar{u}^2 + \frac{2}{3}\bar{u}\bar{v} + \bar{u} \sum_{i+j \geqslant 2} b_{ij}^{(u)} \bar{u}^i \bar{v}^j, \tag{13}$$

$$\bar{\beta}_{\bar{v}} = -\bar{v} + \bar{v}^2 + \frac{12}{8+N}\bar{u}\bar{v} + \bar{v} \sum_{i+j \geqslant 2} b_{ij}^{(v)} \bar{u}^i \bar{v}^j, \tag{14}$$

$$\eta_\phi = \sum_{i+j \geqslant 2} e^{(\phi)}_{ij} \bar{u}^i \bar{v}^j, \tag{15}$$

$$\eta_t = -\frac{2(2+N)}{(8+N)}\bar{u} - \frac{2}{3}\bar{v} + \sum_{i+j \geqslant 2} e^{(t)}_{ij} \bar{u}^i \bar{v}^j, \tag{16}$$

where

$$\bar{\beta}_{\bar{u}} = \frac{3}{16\pi} R_N^{-1} \beta_u, \qquad \bar{\beta}_{\bar{v}} = \frac{3}{16\pi} \beta_v. \tag{17}$$

The coefficients $b^{(u)}_{ij}$, $b^{(v)}_{ij}$, $e^{(\phi)}_{ij}$, and $e^{(t)}_{ij}$ are reported in the tables 1, 2, 3, and 4. Note that due to the rescaling (17), the matrix element of $\Omega$ are two times the derivative of $\bar{\beta}$ with respect to $\bar{u}$ and $\bar{v}$.

**Table 2.** The coefficients $b^{(v)}_{ij}$, cf. equation (14).

| $i,j$ | $(N+8)^i b^{(v)}_{ij}$ |
|---|---|
| 2,0 | $-92.6834 - 5.83417\,N$ |
| 1,1 | $-17.392$ |
| 0,2 | $-0.716174$ |
| 3,0 | $1228.63 + 118.503\,N - 1.83156\,N^2$ |
| 2,1 | $358.882 + 2.84758\,N$ |
| 1,2 | $31.4235$ |
| 0,3 | $0.930766$ |
| 4,0 | $-20723.1 - 2692.\,N - 25.4854\,N^2 - 0.824655\,N^3$ |
| 3,1 | $-8273.27 - 233.78\,N + 0.574757\,N^2$ |
| 2,2 | $-1134.8 - 1.91402\,N$ |
| 1,3 | $-68.4022$ |
| 0,4 | $-1.58239$ |
| 5,0 | $414915. + 67526.8\,N + 1868.92\,N^2 - 13.7618\,N^3 - 0.4602\,N^4$ |
| 4,1 | $211041. + 10633.2\,N - 22.0443\,N^2 + 0.044688\,N^3$ |
| 3,2 | $39732.3 + 365.816\,N - 0.399537\,N^2$ |
| 2,3 | $3666.92 - 0.93257\,N$ |
| 1,4 | $171.066$ |
| 0,5 | $3.26042$ |

We have verified the exactness of our series by the following relations:

(i) $\bar{\beta}_{\bar{u}}(\bar{u},0)$, $\eta_\phi(\bar{u},0)$ and $\eta_t(\bar{u},0)$ reproduce the corresponding functions of the O($N$)-symmetric model [15,21].

(ii) $\bar{\beta}_{\bar{v}}(0,\bar{v})$, $\eta_\phi(0,\bar{v})$ and $\eta_t(0,\bar{v})$ reproduce the corresponding functions of the Ising-like ($N=1$) $\phi^4$ theory.

**Table 3.** The coefficients $e_{ij}^{(\phi)}$, cf. equation (15).

| $i,j$ | $(N+8)^i e_{ij}^{(\phi)}$ |
|---|---|
| 2,0 | $1.83417 + 0.917086\,N$ |
| 1,1 | $0.611391$ |
| 0,2 | $0.0339661$ |
| 3,0 | $-0.873744 - 0.54609\,N - 0.054609\,N^2$ |
| 2,1 | $-0.436872 - 0.054609\,N$ |
| 1,2 | $-0.054609$ |
| 0,3 | $-0.00202255$ |
| 4,0 | $41.5352 + 29.2512\,N + 4.05641\,N^2 - 0.0926845\,N^3$ |
| 3,1 | $27.6901 + 5.65571\,N - 0.123579\,N^2$ |
| 2,2 | $5.40424 + 0.1328\,N$ |
| 1,3 | $0.410151$ |
| 0,4 | $0.0113931$ |
| 5,0 | $-426.896 - 325.329\,N - 57.7615\,N^2 - 1.0524\,N^3 - 0.07092\,N^4$ |
| 4,1 | $-355.747 - 93.2339\,N - 1.5176\,N^2 - 0.1182\,N^3$ |
| 3,2 | $-94.4586 - 5.65221\,N - 0.0262601\,N^2$ |
| 2,3 | $-11.0421 - 0.084259\,N$ |
| 1,4 | $-0.61813$ |
| 0,5 | $-0.0137362$ |

(iii) The following relation holds close to Heisenberg [22,23]

$$\left.\frac{\partial \eta_{\phi,t}}{\partial \overline{v}}\right|_{(\overline{u},0)} = \frac{N+8}{3(N+2)} \left.\frac{\partial \eta_{\phi,t}}{\partial \overline{u}}\right|_{(\overline{u},0)}, \qquad (18)$$

$$\left.\frac{\partial \bar{\beta}_{\overline{v}}}{\partial \overline{v}}\right|_{(\overline{u},0)} + \frac{3(N+2)}{N+8} \left.\frac{\partial \bar{\beta}_{\overline{u}}}{\partial \overline{v}}\right|_{(\overline{u},0)} = \left.\frac{\partial \bar{\beta}_{\overline{u}}}{\partial \overline{u}}\right|_{(\overline{u},0)}, \qquad (19)$$

and Ising fixed points [23]

$$R_N \left( \left.\frac{\partial \bar{\beta}_{\overline{v}}}{\partial \overline{v}}\right|_{(0,\overline{v})} - \left.\frac{\partial \bar{\beta}_{\overline{u}}}{\partial \overline{u}}\right|_{(0,\overline{v})} \right) = \left.\frac{\partial \bar{\beta}_{\overline{v}}}{\partial \overline{u}}\right|_{(0,\overline{v})}, \qquad (20)$$

$$R_N \left.\frac{\partial \eta_\phi}{\partial \overline{v}}\right|_{(0,\overline{v})} = \left.\frac{\partial \eta_\phi}{\partial \overline{u}}\right|_{(0,\overline{v})}. \qquad (21)$$

No analog of such relations exists close to the Ising fixed point for $\eta_t$.

(iv) The following relations hold for $N=1$:

$$\bar{\beta}_{\overline{u}}(u, x-u) + \bar{\beta}_{\overline{v}}(u, x-u) = \bar{\beta}_{\overline{v}}(0, x),$$

**Table 4.** The coefficients $e_{ij}^{(t)}$, cf. equation (16).

| $i,j$ | $(N+8)^i e_{ij}^{(\phi)}$ |
|---|---|
| 2,0 | $13.5025 + 6.751258 N$ |
| 1,1 | $4.50084$ |
| 0,2 | $0.250047$ |
| 3,0 | $-96.7105 - 65.1686\,N - 8.40668\,N^2$ |
| 2,1 | $-48.3553 - 8.40668\,N$ |
| 1,2 | $-6.19023 - 0.116656\,N$ |
| 0,3 | $-0.233588$ |
| 4,0 | $1135.04 + 844.5\,N + 139.656\,N^2 + 0.583377\,N^3$ |
| 3,1 | $756.697 + 184.652\,N + 0.777836\,N^2$ |
| 2,2 | $149.468 + 7.55346\,N$ |
| 1,3 | $11.5154 + 0.115791\,N$ |
| 0,4 | $0.323089$ |
| 5,0 | $-16885.3 - 13691.4\,N - 2885.83\,N^2 - 130.427\,N^3 + 0.14672\,N^4$ |
| 4,1 | $-14071.1 - 4373.98\,N - 217.868\,N^2 + 0.244533\,N^3$ |
| 3,2 | $-3777.55 - 367.372\,N - 2.33704\,N^2$ |
| 2,3 | $-449.218 - 11.589\,N$ |
| 1,4 | $-25.4411 - 0.159321\,N$ |
| 0,5 | $-0.568897$ |

$$\begin{aligned}
\eta_\phi(u, x-u) &= \eta_\phi(0, x), \\
\eta_t(u, x-u) &= \eta_t(0, x).
\end{aligned} \qquad (22)$$

(v) For $N = 2$, one can easily obtain the identities [24,3]

$$\begin{aligned}
\bar\beta_{\bar u}(\bar u + \tfrac{5}{3}\bar v, -\bar v) + \tfrac{5}{3}\bar\beta_{\bar v}(\bar u + \tfrac{5}{3}\bar v, -\bar v) &= \bar\beta_{\bar u}(\bar u, \bar v), \\
\bar\beta_{\bar v}(\bar u + \tfrac{5}{3}\bar v, -\bar v) &= -\bar\beta_{\bar v}(\bar u, \bar v), \\
\eta_\phi(\bar u + \tfrac{5}{3}\bar v, -\bar v) &= \eta_\phi(\bar u, \bar v), \\
\eta_t(\bar u + \tfrac{5}{3}\bar v, -\bar v) &= \eta_t(\bar u, \bar v).
\end{aligned} \qquad (23)$$

These relations are also exactly satisfied by our five-loop series. Note that, since the Ising fixed point is $(0, g_I^*)$, and $g_I^*$ is known with high precision [25]

$$g_I^* = 1.7543637(25), \qquad (24)$$

the above symmetry gives us the location of the cubic fixed point: $((5/3)g_I^*, -g_I^*)$.

(vi) In the large-$N$ limit the critical exponents of the cubic fixed point are related to those of the Ising model: $\eta = \eta_I$ and $\nu = \nu_I$. One can easily see that, for $N \to \infty$, $\eta_\phi(u, v) = \eta_I(v)$, where $\eta_I(v)$ is the perturbative series that determines the exponent $\eta$ of the Ising model. Therefore, the first relation is trivially true. On the other hand, the second relation $\nu = \nu_I$ is not identically satisfied by the series, and is verified only at the critical point [10].

(vii) The series reproduces the previous four-loop results [3].

The obtained RG series are asymptotic and some resummation procedure is needed to extract accurate numerical values for the physical quantities. Exploiting the property of Borel summability of $\phi^4$ theories in two and three dimensions, we resum the divergent asymptotic series by a Borel transformation combined with a method for the analytic extension of the Borel transform. This last procedure can be obtained by a Padé extension or by a conformal mapping [26] which maps the domain of analyticity of the Borel transform onto a circle (see [19,26] for details).

The conformal mapping method takes advantage of the knowledge of the large order behavior of the perturbative series $F(\overline{u}, z) = \sum_k f_k(z) \overline{u}^k$ [3,10]

$$f_k(z) \sim k! \, (-a(z))^k \, k^b \left[1 + O(k^{-1})\right] \qquad \text{with} \qquad a(z) = -1/\overline{u}_b(z), \qquad (25)$$

where $\overline{u}_b(z)$ is the singularity of the Borel transform closest to the origin at fixed $z = \bar{v}/\bar{u}$, given by [3]

$$\frac{1}{\overline{u}_b(z)} = -a\,(R_N + z) \quad \text{for} \quad 0 < z, \qquad (26)$$

$$\frac{1}{\overline{u}_b(z)} = -a\left(R_N + \frac{1}{N}z\right) \quad \text{for} \quad 0 > z > -\frac{2NR_N}{N+1},$$

where $a = 0.238659217\ldots$

It should be noted that these results do not apply to the case $N = 0$. Indeed, in this case, additional singularities in the Borel transform are expected [27].

For each perturbative series $R(\bar{u}, \bar{v})$, we obtain the following approximants

$$E(R)_p(\alpha, b; \overline{u}, \overline{v}) = \sum_{k=0}^{p} B_k(\alpha, b; \overline{v}/\overline{u}) \times \int_0^\infty dt\, t^b e^{-t} \frac{y(\overline{u}t; \overline{v}/\overline{u})^k}{[1 - y(\overline{u}t; \overline{v}/\overline{u})]^\alpha}, \qquad (27)$$

where

$$y(x; z) = \frac{\sqrt{1 - x/\overline{u}_b(z)} - 1}{\sqrt{1 - x/\overline{u}_b(z)} + 1}, \qquad (28)$$

and the coefficients $B_k$ are determined by the condition that the expansion of $E(R)_p(\alpha, b; \overline{u}, \overline{v})$ in powers of $\overline{u}$ and $\overline{v}$ gives $R(\overline{u}, \overline{v})$ to order $p$.

Within the second resummation procedure, the Borel-Leroy transform is analytically extended by means of a generalized Padé approximant technique, using the resolvent series trick (see, e. g. [9]). Explicitly, once introduced the resolvent series of the perturbative one $R(\bar{u}, \bar{v})$

$$\tilde{P}(R)(\overline{u}, \overline{v}, b, \lambda) = \sum_n \lambda^n \sum_{k=0}^{n} \frac{\tilde{P}_{k,n}}{(n+b)!} \overline{u}^{n-k} \overline{v}^k, \qquad (29)$$

which is a series in powers of $\lambda$ with coefficients being uniform polynomials in $\overline{u}, \overline{v}$. The analytical continuation of the Borel transform is the Padè approximant $[N/M]$ in $\lambda$ at $\lambda = 1$. Obviously, the approximant for each perturbative series depends on the chosen Padé approximant and on the parameter $b$.

An important issue in the fixed dimension approach to critical phenomena (and in general of all the field theoretical methods) concerns the analytic properties of the $\beta$-functions. As shown in [16] for the $O(N)$ model, the presence of confluent singularities in the zero of the perturbative $\beta$ function causes a slow convergence of the results given by the resummation of the perturbative series to the correct fixed point value. The $O(N)$ two-dimensional field-theory estimates of physical quantities [15,26] are less accurate than the three-dimensional ones due to the stronger non-analyticities at the fixed point [13,14,16], to say nothing about the stronger growth of the series coefficients themselves. In [16] it is shown that the nonanalytic terms may cause large imprecisions in the estimate of the exponent related to the leading correction to the scaling $\omega$. At the same time, the result for the fixed point location turns out to be a rather good approximation of the accurate one.

## 3. Analysis of five loop series

### 3.1. Stability of $O(N)$ and Ising fixed points

First of all, we analyze the stability properties of the $O(N)$-symmetric fixed point. Since $\partial_{\overline{u}}\overline{\beta}_{\overline{v}}|_{(\overline{u},0)} = 0$, the stability of the fixed point with respect to an anisotropic cubic perturbation is given by

$$\omega_2 = 2 \frac{\partial \bar{\beta}_{\overline{v}}}{\partial \overline{v}}(\overline{u}^*, 0), \tag{30}$$

where $\overline{u}^*$ is the fixed-point value of the $O(N)$ vector model. In table 5 we report the results for $\omega_2$ for several values of $N$. The $O(N)$ fixed point results unstable for $N \geqslant 3$. For $N = 2$ our result is compatible with $\omega_2 = 0$, which is essential in the context of a continuous line of the fixed points expected.

**Table 5.** Half of the exponent $\omega_2$ at the $O(N)$ fixed point. CM is the value obtained using conformal mapping technique and PB the one using Padé-Borel.

| N | CM 4-loop | PB 4-loop | CM5-loop | PB 5-loop |
|---|-----------|-----------|-----------|-----------|
| 2 | 0.03(3)   | 0.06(4)   | 0.025(40) | 0.00(5)   |
| 3 | −0.08(3)  | −0.07(3)  | −0.10(6)  | −0.10(5)  |
| 4 | −0.18(4)  | −0.17(5)  | −0.20(3)  | −0.22(4)  |
| 8 | −0.45(5)  | −0.44(6)  | −0.48(4)  | −0.50(5)  |

Then we focus our attention on the stability properties of the Ising fixed point. Also in this case the stability is given by

$$\omega_2 = 2 \frac{\partial \bar{\beta}_{\overline{u}}}{\partial \overline{u}}(0, \overline{v}^*), \tag{31}$$

where $\overline{v}^*$ is the fixed-point value of the Ising model (24). As expected, the series $\omega_2(\overline{v})$ is independent of $N$

$$\frac{\omega_2^I(\overline{v})}{2} = -1 + \frac{2}{3}\overline{v} - 0.2161\overline{v}^2 + 0.23157\overline{v}^3 - 0.31169\overline{v}^4 + 0.555161\overline{v}^5. \quad (32)$$

The fixed point value of this exponent is $\omega_2^I/2 = -0.09(8)$, using the conformal mapping method, and $-0.08(10)$ using the Padé-Borel analysis. We note that the [4/1] approximant with $b = 1$ leads to $\omega_2^I/2 = -0.031$. These values are compatible with the exact known result $\alpha_I = 0$.

### 3.2. Analysis for $N \geqslant 3$

We first analyze the position of the cubic fixed point for $N \geqslant 3$, previously found in the four-loop approximation [3]. It was claimed in [3] that the quite peculiar features of this fixed point (like the marginal instability) make its existence quite doubtful and that it might be an artifact of the *relatively* short series available at that time. Now we are in position to analyze longer series and to further confirm or to reject this statement.

**Table 6.** Coordinates of the cubic fixed point in the 4-loop and 5-loop approximations.

| N | 4-loop | 5-loop | Padé [4/1] b=1 |
|---|--------|--------|----------------|
| 3 | [0.83(12),1.12(9)] | [0.54(6),1.35(4)] | [0.050,1.757] |
| 4 | [0.54(10),1.43(8)] | [0.32(5),1.58(4)] | [0.031,1.774] |
| 8 | [0.24(8),1.72(10)] | [0.14(4),1.74(4)] | [0.015,1.788] |

The results obtained with the conformal mapping methods are reported in table 6 together with the old four loop results, in order to make the comparison visible at first sight. This table shows that the position of the cubic fixed point drastically moves close to the Ising fixed point with increasing the order of perturbation theory from four to five loops. Both at four and at five loops, the quoted errors are less than the difference between the two estimates, leading to the conclusion that the reported uncertainty is actually an underestimate of the correct one. We remind to the reader that this error comes from the so-called stability criterion, i. e. it is obtained looking at those approximants that minimize the difference between the estimates at the two highest available orders. These considerable discrepancies between the four- and five-loop results lead to serious doubts in the existence of the cubic fixed point in two dimensions.

In order to better understand these strange results, we report the values of the coordinate $u^*$ obtained for the cubic fixed point using several Padé approximants for $N = 0, 3, 4, 8$; the estimates are presented in figure 1 as functions of $b$. Let us consider first the case $N = 3$ as a typical example. If one limits himself with only three lower-order approximants [2/1], [3/1], and [2/2], he easily finds that they minimize their differences under $b \sim 0$, leading in such a way to the estimate $\overline{u}^* \sim 0.7$. Taking into account two non-defective Padé approximants [4/1] and [3/2] (note the oscillating behavior of the [3/2] approximant for $b < 1$, signaling the presence of close singularities) existing at the five-loop level shifts the zone of stability to $b \sim 2$, thus leading to the estimate $\overline{u}^* \sim 0.5$. Moreover, the approximant [4/1] with $b = 1$, that is usually considered as one of the best approximants, results in the estimate $\overline{u}^* = 0.050$ very close to zero. Because of the alternative character of RG expansions, it looks very likely that the unknown six-loop contribution (and the higher-order ones) will locate the stability region somewhere near $b \sim 1$, leading finally to the coalescence of the Ising and cubic fixed point. According to this scenario, the cubic fixed point, found at finite order in perturbation theory, is probably only an artifact due to the finiteness of the perturbative series.

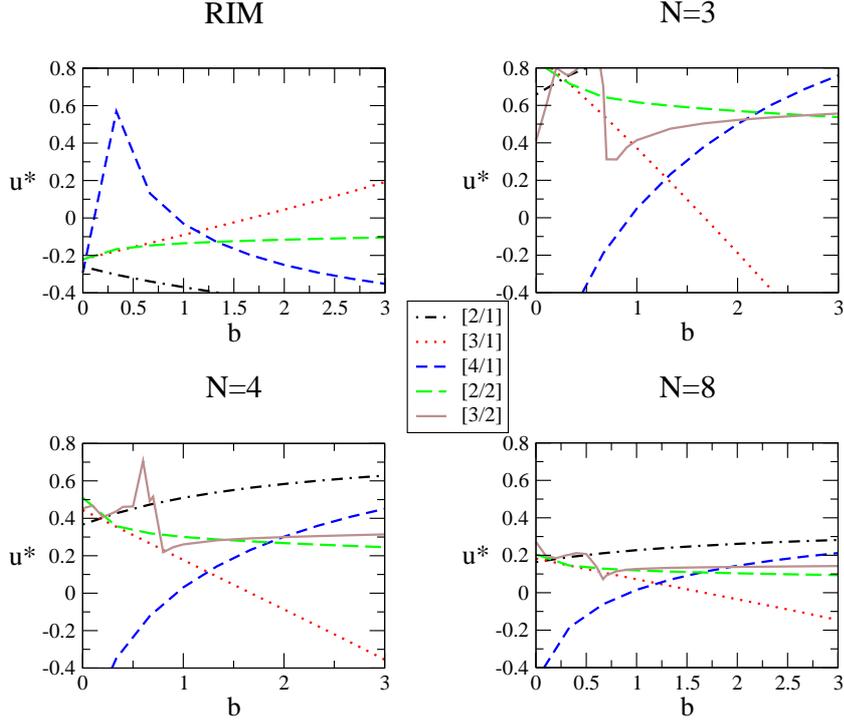

**Figure 1.** Coordinate $u^*$ of the cubic fixed point within Padé-Borel method for some values of N.

The same scenario is also possible for other values of $N$. From figure 1 we see that the region of maximum stability always shifts from $b \sim 0$ to $b \sim 2$ with increasing the order of approximation from four to five loops, moving the coordinate of the fixed point from $\overline{u}^* \sim 0.5$ for $N = 4$ ($\overline{u}^* \sim 0.2$ for $N = 8$) to $\overline{u}^* \sim 0.3$ ($\overline{u}^* \sim 0.1$). Let us stress again that the value given by the approximant [4/1] with $b = 1$ is always very close to zero, as is seen from table 6. Note that, with increasing N, the distance of the cubic fixed point from the Ising one reduces fastly.

In the limit $N \to \infty$ the series simplify as at the four-loop level (see [3]). We only mention that with increasing the length of the RG series the coordinate $\overline{u}^*$ of the cubic fixed point shifts from $\overline{u}^* \sim 0.08$ to $\overline{u}^* \sim 0.03$ that again is much closer to zero.

Finally, we briefly discuss the fate of the random fixed point governing the critical behavior of the weakly-disordered Ising model which is described by the Hamiltonian (1) in the replica limit $N \to 0$. We do not use here advanced resummation procedures developed [28] to avoid Borel non-summability at fixed $u/v$ [27], but limit ourselves by a simple Padé analysis, since it is sufficient for our aims. We find, for the majority of the considered approximants, a fixed point with negative $\overline{u}^*$. A possible estimate, according to stability criteria is $\overline{u}^* = -0.1(1)$, but if we concentrate on some certain approximants we obtain $\overline{u}^* = -0.090$ for the [3/1] and $\overline{u}^* = -0.030$ for the [4/1] (both with $b = 1$). In particular, the last value is very close to zero, i. e. to the value predicted by the asymptotically exact solution that has been obtained in the framework of the fermionic representation [2,5,6]. It is worth noting that the five-loop results for $N = 0$ seem to be less scattered than analogous four-loop estimates obtained by means of Chisholm-Borel resummation technique [29], and they look more precise than their five-loop counterparts for finite $N$.

### 3.3. Analysis for $N = 2$

The four-loop analysis of [3] for $N = 2$ turned out to be compatible with the presence of a line of the fixed points joining the $O(2)$-symmetric and the decoupled Ising fixed points. The lines of zeros of the two $\beta$ functions were found to be practically parallel and the quoted error was bigger than the distance between them. This line of fixed points with continuously varying critical exponents is in agreement with what is expected from the correspondence, at the critical point, between the cubic model and the Ashkin-Teller and the planar model with fourth-order anisotropy [1,3]. We are now in a position to verify this statement at the five-loop level.

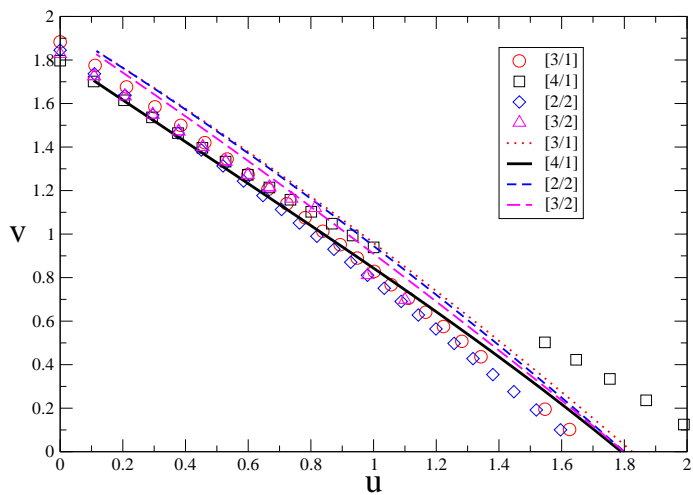

**Figure 2.** Zeros of $\beta_u$ (continuous lines) and $\beta_v$ (points) for several Padé approximants (all with $b = 1$).

First, we analyze the series with the conformal mapping method. Again we find that zeros of two $\beta$ functions form two parallel lines, while the apparent uncertainties seem to be smaller than their separation. Of course, this fact may simply indicate that the model has no fixed point at all. Let us, however, look more attentively at this result and, in particular, at the accuracy of the quoted error. In fact, as we have already seen for $N > 2$, the error coming from stability criteria is likely an underestimate of the correct one. To better understand the situation, we use the Padé-Borel method. In figure 2 we report the curves of zeros of the two $\beta$ functions given by several Padé approximants under $b = 1$, the value that for $N > 2$ always leads to good results and that is the best for the fixed point values of $O(N)$ and Ising model [15]. The four approximants for $\beta_u$ are always well-defined. They are hardly distinguishable close to the $O(2)$ fixed point and their separation slowly increases moving toward the $v$-axis. The coordinate of the Ising fixed point $\overline{v} \sim 1.8$ is obtained using the [4/1] approximant, since approximants of [L-1/1] type proved to give rather precise estimates for a fixed point location both in two and three [30,31] dimensions. The situation is a bit worse for $\beta_v$ function. In fact, the working approximants are well defined close to the Ising fixed point, but approaching the $\overline{u}$ axis they become defective. The approximant [3/2] starts oscillating around $\overline{u} \sim 0.8$, while [4/1] is bad in the range $\overline{u} \in [1, 1.5]$ and [3/1] for $\overline{u} > 1.3$. Also the values of zeros of the approximant [4/1] for $\overline{u} > 1.5$ are not reliable enough since they may suffer from the effect of close singularities.

Despite these shortcomings, we can obtain a rich information from figure 2. Indeed, the line of zeros of $\beta_u$ given by the approximant [4/1] practically coincides with those of the $\beta_v$ from the Ising fixed point up to $\overline{u} \sim 0.8$. For greater $\overline{u}$, various approximants for $\beta_v$ result in the lines of zeros that diverge leaving, however, the line [4/1] of $\beta_u$ zeros between them. Keeping in mind a finite length of the RG series and the effect of non-analytic terms missed by the perturbation theory, we retain this fact as a strong evidence in favor of the continuous line of fixed points. The best estimate for this line is believed to be that given by the approximant [4/1] for $\beta_u$. Thus it will be used in what follows to calculate continuous varying critical exponents.

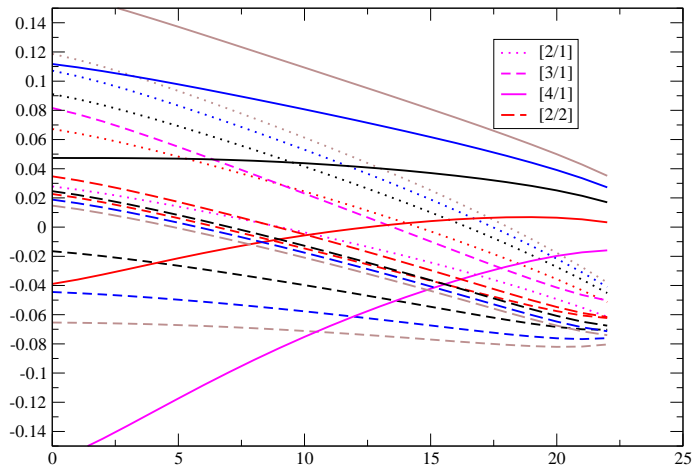

**Figure 3.** Smallest eigenvalues of the $\Omega$ matrix along the line of fixed points for several Padé approximant. The various curves corresponding to the same approximant correspond to different choices of $b$ from 0 to 1.

First, we check that the smallest eigenvalues of the $\Omega$ matrix are compatible with zero, which is a necessary condition for getting a line of fixed points. Some Padé-Borel results are shown in figure 3. The figure seems somewhat complicated. However, one can realize at first glance that the smallest eigenvalue is always very close to zero.

When evaluating the critical exponents, one should keep in mind that the limit $z \to 0$ is not simply accessible perturbatively since it corresponds to the 2D XY model which is known to behave in a quite specific manner. In particular, it does not undergo an ordinary transition into the ordered phase at any finite temperature and its critical behavior is essentially controlled by the vortex excitations. Such excitations lead to an exponentially diverging correlation length at finite temperature. This behavior cannot be accounted for within the $\lambda\phi^4$ model equation (1) dealt with in this paper. Since a new physics emerges after arriving at the point $z = 0$, it is natural to assume that the observables as functions of the fixed point location may be non-analytic near this point. Hence, what we really can explore trusting upon our five-loop expansions is a domain corresponding to finite (and not too small) values of $z$. Oppositely, the limit $z \to \infty$ (or $1/z \to 0$) looks not at all dangerous in the above sense since it corresponds to the critical behavior close to that of the Ising model, which was shown not to be considerably effected by the non-analytic terms even in two dimensions [15]. Note that the above presented results concerning the line of the fixed point confirm this idea. Indeed, as seen from figure 2, at the "Ising side", i. e. for $0 < \overline{u} \sim 0.8$, the zeros of both $\beta$ functions form smooth curves running very close to each other. However, the closer the "XY side" is the stronger the estimates for $\beta$ function zeros are scattered, probably indicating the increasing impact of non-analytic contributions.

The expected value of $\eta$ is 1/4 independent of the location of a fixed point within the line. A clever way to check the constantness of $\eta$ along the line of fixed points is probably to write the RG function in terms of $\overline{u}$ and $x = \overline{u} + \overline{v}$, i.e. $\eta_x(x, \overline{u}) = \eta(\overline{u}, x - \overline{u})$. Then one resums the difference $\Delta(x, \overline{u}) = \eta_x(x, \overline{u}) - \eta_x(x, 0)$. Along the line for all the five-loop approximants we always find $|\Delta(x, \overline{u})| < 8 \times 10^{-3}$. This leads us to conclude that the two-dimensional LGW approach is capable of keeping the constantness of $\eta$ within an error of 3%. Note that the previous quoted problems about non-analyticities close to the $O(2)$ side do not significantly affect the estimates of $\eta$

For the exponent $y = \eta - \eta_t = 2 - 1/\nu$ it was conjectured in [3] that it should behave like
$$y = \frac{2}{1+x}, \qquad \text{where,} \qquad x = \frac{2}{\pi} \arctan \frac{\overline{v}^*}{\overline{u}^*}. \qquad (33)$$
A direct reliable quantitative estimate of this exponent is impossible due to the effect of nonanalyticities, in particular close to the $O(2)$ fixed point for the reasons explained above. In fact, we know from [15] that the resummation of $y$ at the Ising fixed point $y$ provides 1.04, very close to the exact value 1. Instead, at the $O(2)$ fixed point one has $y \sim 1.25$, which is quite far from a diverging $\nu$, i.e. $y = 2$. A direct resummation of this exponent is reported in figure 4, and as expected it seems to reproduce the correct critical behavior only close to the Ising fixed point.

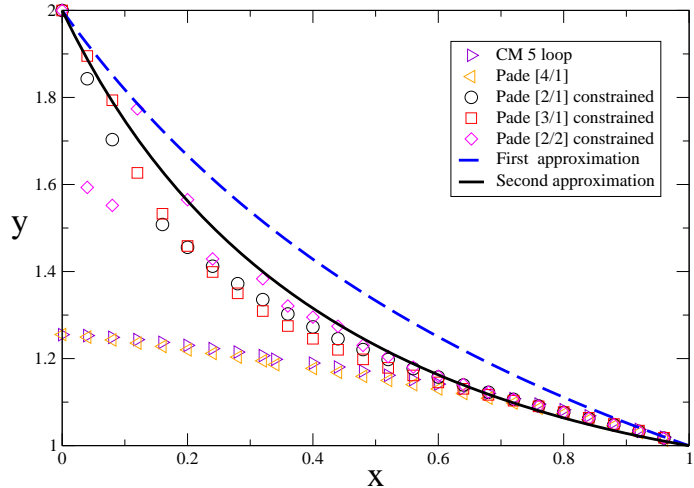

**Figure 4.** The exponent $y$ as function of $x$ from free and constrained resummation. The solid line is the conjecture equation (34).

In [3] it was proposed to use the method of constraining the exponents to assume the exactly known value along the axes in order to get better quantitative results. We apply here a quite different method of constrained analysis with respect to [3]. Without going into technical details we have preferred to constrain the series of $y$ expressed in terms of $\overline{u}$ and $x = \overline{u} + \overline{v}$ (as previously for the exponent $\eta$). The results obtained by Padé-Borel method are again reported in figure 4. They are practically indistinguishable from the unconstrained ones up to $x \sim 0.6$, and then they start oscillating, signaling the presence of singularities and of bad quantitative estimates. The conformal mapping results are practically equivalent. The numerical data thus obtained are well fitted by the conjectured curve:

$$y = \frac{4}{4 - (1-x)(2-x)}, \qquad (34)$$

but we are not able to estimate the goodness of our resummation and, as a result, to verify the new conjecture equation (34). Perhaps, the exact behavior of the exponent $\nu$ along the line of fixed points requires a new method of analysis of the perturbative series and, in any case, it deserves additional studies on the subject like Monte Carlo simulation or high temperature expansion.

## 4. Conclusions

To summarize, the critical behavior of the two-dimensional cubic model with the $N$-vector order parameter has been studied. The five-loop contributions to the $\beta$ functions and critical exponents have been calculated and the five-loop RG series have been resummed by means of Padé-Borel-Leroy procedure and using the conformal mapping technique. For $N = 2$ we have found that the continuous line of fixed points connecting the Heisenberg and the Ising ones is well reproduced by the resummed five-loop RG series. Moreover, the five-loop terms make the lines of zeros of $\beta$ functions for $\overline{u}$ and $\overline{v}$ closer to each other, thus improving the results of the lower-order approximation. For $N > 2$, the five-loop contributions have been shown to shift the cubic fixed point, given by the four-loop approximation, towards the Ising fixed point. This may be considered as an argument in favor of the idea that the existence of cubic fixed point in two dimensions for $N \geqslant 3$ is an artifact of the perturbative analysis. The model with $N = 0$ describing the critical thermodynamics of 2D weakly-disordered Ising systems has been also studied. The results obtained have been found to be compatible with the conclusion that in two dimensions the impure critical behavior is governed by the Ising fixed point.

## Acknowledgements

We are grateful to Pietro Parruccini and Ettore Vicari for discussions. The authors acknowledge the financial support of EPSRC under Grant No. GR/R83712/01 (P.C.), the Russian Foundation for Basic Research under Grant No. 04–02–16189 (A.I.S., E.V.O., D.V.P.), and the Ministry of Education of Russian Federation under Grant No. E02–3.2–266 (A.I.S., E.V.O., D.V.P.). A.I.S. has much benefitted from the warm hospitality of Scuola Normale Superiore and Dipartimento di Fisica dell'Universitá di Pisa, where the major part of this research was done.

## References

1. José J.V., Kadanoff L.P., Kirkpatrick S., Nelson D.R., Phys. Rev. B, 1977, **16**, No. 3, 1217–1241.
2. Shalaev B.N., Phys. Reports, 1994, **237**, No. 3, 129–188.
3. Calabrese P., Celi A., Phys. Rev. B, 2002, **66**, No. 18, 184410.
4. Shalaev B.N., Fiz. Tverd. Tela, 1989, **31**, No. 1, 93–101 (in Russian);
   Sov. Phys. Solid State, 1989, **31**, No. 1, 51.
5. Dotsenko Vl.S., Dotsenko Vik.S., Adv. Phys., 1983, **32**, 129.
6. Shalaev B.N., Fiz. Tverd. Tela, 1984, **26**, No. 10, 3002–3005 (in Russian);
   Sov. Phys. Solid State, 1984, **26**, No. 10, 1811–1813.
7. Kleinert H., Schulte-Frohlinde V., Phys. Lett. B, 1995, **342**, No. 1–4, 284–296;
   Kleinert H., Thoms S., Phys. Rev. D, 1995, **52**, No. 10, 5926–5943;
   Kleinert H., Thoms S., Schulte-Frohlinde V., Phys. Rev. B, 1997, **56**, No. 22, 14428–14434.
8. Shalaev B.N., Antonenko S.A., Sokolov A.I., Phys. Lett. A, 1997, **230**, No. 1–2, 105–110.
9. Pakhnin D.V., Sokolov A.I., Phys. Rev. B, 2000, **61**, No. 22, 15130–15135.
10. Carmona J.M., Pelissetto A., Vicari E., Phys. Rev. B, 2000, **61**, No. 22, 15136–15151.
11. Folk R., Holovatch Yu., Yavors'kii T., Phys. Rev. B, 2000, **62**, No. 18, 12195–12200;
12. Pakhnin D.V., Sokolov A.I., Phys. Rev. B, 2001, **64**, No. 9, 094407.
13. Nickel B.G., Phase Transitions, ed. by Lévy M., Le Guillou J.C., Zinn-Justin J. Plenum, New York-London, 1982;
    Nickel B.G., Physica A, 1991, **177**, No. 1, 189–196.

14. Pelissetto A., Vicari E., Nucl. Phys. B, 1998, **519**, No. 3, 626–660;
    Nucl. Phys. B (Proc. Suppl.), 1999, **S73**, 775–777.
15. Orlov E.V., Sokolov A.I., Fiz. Tverd. Tela, 2000, **42**, No. 11, 2087–2093 (in Russian);
    Phys. Solid State, 2000, **42**, No. 11, 2151–2158.
16. Calabrese P., Caselle M., Celi A., Pelissetto A., Vicari E., J. Phys. A, 2000, **33**, No. 46, 8155–8170.
17. Aharony A., Phase Transitions and Critical Phenomena, ed. by Domb C., Lebowitz J, vol. 6, p. 357. Academic Press, New York, 1976.
18. Pelissetto A., Vicari E., Phys. Reports, 2002, **368**, No. 6, 549–727.
19. Zinn-Justin J. Quantum Field Theory and Critical Phenomena, fourth edition. Clarendon Press, Oxford, 2002.
20. Calabrese P., Pelissetto A., Rossi P., Vicari E., Int. J. Mod. Phys. B, 2003, **17**, No. 31–32, 5829–5838.
21. Unfortunately, the five-loop contribution to the critical exponent $\gamma$ ($\eta_t$) reported in [15] contains numerical errors. Here, the correct values of corresponding coefficients are presented (table 4).
22. Calabrese P., Pelissetto A., Vicari E., Phys. Rev. B, 2003, **67**, No. 2, 024418;
    Acta Phys. Slov., 2002, **52**, 311–316.
23. Calabrese P., Parruccini P., Pelissetto A., Vicari E., in preparation.
24. Korzhenevskii A.L., Zh. Eksp. Teor. Fiz., 1976, **71**, No. 4, 1434–1442 (in Russian);
    Sov. Phys. JETP, 1976, **44**, No. 4, 751–757.
25. Caselle M., Hasenbusch M., Pelissetto A., Vicari E., J. Phys. A, 2000, **33**, No. 46, 8171–8180;
    J. Phys. A, 2001, **34**, No. 14, 2923–2948.
26. Le Guillou J. C., Zinn-Justin J., Phys. Rev. Lett., 1977, **39**, No. 2, 95–98;
    Phys. Rev. B, 1980, **21**, No. 9, 3976–3998.
27. Bray A.J., McCarthy T., Moore M.A., Reger J.D., Young A.P., Phys. Rev. B, 1987, **36**, No. 4, 2212–2219;
    McKane A.J., Phys. Rev. B, 1994, **49**, No. 17, 12003–12009;
    Alvarez G., Martin-Mayor V., Ruiz-Lorenzo J.J., J. Phys. A, 2000, **33**, No. 5, 841–850.
28. Pelissetto A., Vicari E., Phys. Rev. B, 2000, **62**, No. 10, 6393–6409.
29. Mayer I.O., Sokolov A.I., Shalaev B.N., Ferroelectrics, 1989, **95**, No. 1, 93–96.
30. Antonenko S.A., Sokolov A.I., Phys. Rev. E, 1995, **51**, No. 3, 1894–1898.
31. Sokolov A.I., Fiz. Tverd. Tela, 1998, **40**, No. 7, 1284–1290 (in Russian);
    Phys. Solid State, 1998, **40**, No. 7, 1169–1174.